\documentclass{emulateapj}

\usepackage{float}
\usepackage{amsmath}
\usepackage{epsfig,floatflt}
\usepackage{subfigure}

\begin{document}

\title{A re-analysis of the three-year \emph{Wilkinson Microwave Anisotropy
    Probe} temperature power spectrum and likelihood}

\author{H.\ K.\ Eriksen\altaffilmark{1,2,3,4,5}, Greg Huey\altaffilmark{6},
  R. Saha\altaffilmark{7}, F. K. Hansen\altaffilmark{2,3}, J.
  Dick\altaffilmark{8}, A. J. Banday\altaffilmark{9}, K. M.
  G\'{o}rski\altaffilmark{4,5,10}, P. Jain\altaffilmark{7}, J. B.
  Jewell\altaffilmark{4}, L.  Knox\altaffilmark{8}, D. L.
  Larson\altaffilmark{11}, I.\ J.\ O'Dwyer\altaffilmark{4,5}, T.
  Souradeep\altaffilmark{12}, B.\ D.\ Wandelt\altaffilmark{6,11}}

\altaffiltext{1}{email: h.k.k.eriksen@astro.uio.no}

\altaffiltext{2}{Institute of Theoretical Astrophysics, University of
Oslo, P.O.\ Box 1029 Blindern, N-0315 Oslo, Norway}

\altaffiltext{3}{Centre of
Mathematics for Applications, University of Oslo, P.O.\ Box 1053
Blindern, N-0316 Oslo}

\altaffiltext{4}{Jet Propulsion Laboratory, 4800 Oak
  Grove Drive, Pasadena CA 91109} 

\altaffiltext{5}{California Institute of Technology, Pasadena, CA
  91125} 

\altaffiltext{6}{Department of Physics, University of Illinois,
  Urbana, IL 61801}

\altaffiltext{7}{Physics Department, Indian Institute of Technology,
  Kanpur, U.P., 208016, India}

\altaffiltext{8}{Department of Physics, One Shields Avenue, University
  of California, Davis, California 95616}

\altaffiltext{9}{Max-Planck-Institut f\"ur Astrophysik,
Karl-Schwarzschild-Str.\ 1, Postfach 1317, D-85741 Garching bei
M\"unchen, Germany}

\altaffiltext{10}{Warsaw University Observatory, Aleje Ujazdowskie 4, 00-478 Warszawa,
  Poland}

\altaffiltext{11}{Astronomy Department, University of Illinois at
  Urbana-Champaign, IL 61801-3080}

\altaffiltext{12}{Inter-University Centre for Astronomy and
  Astrophysics, Post Bag 4, Ganeshkhind, Pune 411007, India}

\date{Received - / Accepted -}

\begin{abstract}
  We analyze the three-year \emph{WMAP} temperature anisotropy data
  seeking to confirm the power spectrum and likelihoods published by
  the \emph{WMAP} team.  We apply five independent implementations of
  four algorithms to the power spectrum estimation and two
  implementations to the parameter estimation.  Our single most
  important result is that we broadly confirm the \emph{WMAP} power
  spectrum and analysis. Still, we do find two small but potentially
  important discrepancies: On large angular scales there is a small
  power excess in the \emph{WMAP} spectrum (5--10\% at $\ell \lesssim
  30$) primarily due to likelihood approximation issues between $13
  \le \ell \lesssim 30$. On small angular scales there is a systematic
  difference between the V- and W-band spectra (few percent at $\ell
  \gtrsim 300$). Recently, the latter discrepancy was explained by
  \citet{huffenberger:2006} in terms of over-subtraction of unresolved
  point sources.  As far as the low-$\ell$ bias is concerned, most
  parameters are affected by a few tenths of a sigma. The most
  important effect is seen in $n_{\textrm{s}}$. For the combination of
  \emph{WMAP}, Acbar and BOOMERanG, the significance of
  $n_{\textrm{s}} \ne 1$ drops from $\sim2.7\sigma$ to $\sim2.3\sigma$
  when correcting for this bias.  We propose a few simple improvements
  to the low-$\ell$ \emph{WMAP} likelihood code, and introduce two
  important extensions to the Gibbs sampling method that allows for
  proper sampling of the low signal-to-noise regime. Finally, we make
  the products from the Gibbs sampling analysis publically available,
  thereby providing a fast and simple route to the exact likelihood
  without the need of expensive matrix inversions.
\end{abstract}


\maketitle

\section{Introduction}

Very recently, the \emph{Wilkinson Microwave Anisotropy Probe}
(\emph{WMAP}) team released the results from three years of
observations of the cosmic microwave background \citep{hinshaw:2006,
page:2006, spergel:2006}. In addition to improving upon the successful
first-year release \citep{bennett:2003a}, this new data set includes
the first full-sky polarization maps at millimeter wavelengths
\citep{page:2006}, and is destined to be of great importance for the
CMB community for many years to come.

Given the dominant position of the \emph{WMAP} experiment in our
current understanding of cosmology, it is imperative that all of the
results are verified by several external groups using independent
techniques.  In this paper we begin these efforts by re-computing the
angular power spectrum and likelihoods\footnote{We use version
v2p1 of the \emph{WMAP} likelihood in this paper;
http://lambda.gsfc.nasa.gov/}, arguably the two most important
products of any CMB experiment. Similar work was carried out after the
first-year \emph{WMAP} release in 2003 by several groups (e.g.,
Slosar, Seljak and Makarov, 2004 -- the importance of foreground
marginalization at large scales; O'Dwyer et al.\ 2004 -- the
importance of exact likelihood/posterior evaluation at large scales;
Fosalba and Szapudi 2004 -- pixel space versus harmonic space
computations).  These analyses significantly contributed to the
improvements that were implemented for the three-year \emph{WMAP} data
release \citep{hinshaw:2006}.

Our philosophy in the present paper is that of multiple redundancy and
cross-checking: For both the angular power spectrum and the
likelihoods (or posteriors) we apply several different algorithms and
in some cases different implementations of the same algorithm.
Additionally, the data were downloaded, prepared and analyzed by five
independent groups, further reducing the risk of both programming
errors, simple misunderstandings and algorithmic peculiarities. This
increases our confidence in the final results.

One question that will be given particular attention is the issue of
statistical and numerical procedure. Examples are, how should a high
signal-to-noise but low-$\ell$ likelihood be regularized at its small
scale cut-off in order to avoid numerical artifacts? And in what
$\ell$-range does a pseudo-spectrum estimator provide an adequate
approximation to the full posterior, and where should an exact
approach (such as Gibbs sampling) be preferred?


A separate question is the problem of foreground modeling and
subtraction. This will be considered in greater detail in a future
publication, and for now we mainly adopt the approaches used by the
\emph{WMAP} team, with one or two notable exceptions: A power spectrum
that does not rely on external information \citep{saha:2006} is
established and used as a cross-check on the results that are obtained
from template-corrected maps, and we compare spectra computed for
different frequencies in order to search for frequency-dependent
signatures.

Another major part of the three-year \emph{WMAP} release are
measurements of the CMB polarization \citep{page:2006}. This is a very
complex data set, and requires a high degree of algorithmic
sophistication for proper interpretation. The necessary extensions to
the methods described in the present paper are currently being
developed, and the results from the corresponding analyses will be
reported upon as soon as they are completed. In the present paper, we
focus on temperature information only.

The rest of the paper is organized as follows. In Section
\ref{sec:methods} we briefly list the methods we use, and in Section
\ref{sec:data} we describe the data. Then, we focus on the low-$\ell$
part of the power spectrum in Section \ref{sec:large_scale} and the
high-$\ell$ part in Section \ref{sec:small_scale}. Cosmological
parameters are considered in Section \ref{sec:parameters}, before
drawing conclusions in Section \ref{sec:conclusions}. Algorithmic
details are deferred to two Appendices.

\section{Methods}
\label{sec:methods}

In this paper we are primarily interested in scientific results, and
less in algorithms as such. In this section we only list each method
that we use, and provide more details in the Appendices where
necessary, or with references to earlier papers.

\paragraph{Low-$\ell$ likelihood evaluation}

A central component for several of our methods is evaluation of the
full likelihood $\mathcal{L}$ in pixel-space which is defined by
\begin{equation}
-2\log \mathcal{L} = \mathbf{d}^{t} \mathbf{C}^{-1} \mathbf{d} + \log
|\mathbf{C}|.
\label{eq:lnl_def}
\end{equation}
up to a normalization constant. Here $\mathbf{d}$ are the observed
data in a pixelized form and $\mathbf{C}$ is the corresponding
covariance matrix. Evaluation of this quantity involves computing a
matrix inverse and determinant, and therefore scales as
$\mathcal{O}(N_{\textrm{pix}}^3)$, $N_{\textrm{pix}}$ being the number
of pixels in the data set.  Consequently, likelihood evaluations are
only feasible at low resolutions, and the data must be degraded prior
to analysis. Full details on this operation are given in Appendix
\ref{app:lowl_likelihood}.  Regularization of the covariance matrix
with respect to the Nyquist frequency is an issue worth some
consideration and we do this by wide-beam smoothing and noise
addition.  This is to be contrasted to the approach used by the
\emph{WMAP} likelihood code that goes beyond the Nyquist frequency to
achieve a similar effect \citep{hinshaw:2006}.  For further comments
on this issue, see Section \ref{sec:large_scale} and Appendix
\ref{app:lowl_likelihood}.

\paragraph{Maximum--likelihood estimation}

The maximum-likelihood (ML) power spectrum is simply the one that
maximizes the likelihood as defined by Equation \ref{eq:lnl_def}. This
may be found by any non-linear search algorithm, and for the present
paper we have used both a quasi-Newton and a Sequential Quadratic
Programming algorithm for this task.  We obtain identical results with
the two methods. Previous analyses using similar techniques include
\citet{gorski:1996}, \cite{bond:1998}, \citet{efstathiou:2004},
\citet{slosar:2004} and \citet{hinshaw:2006}.

\paragraph{Posterior mapping by MCMC}

Alternatively, we may choose a Bayesian approach and map out the
entire posterior distribution $P(C_{\ell} | \mathbf{d}) \propto
\mathcal{L}(C_{\ell}) P(C_{\ell})$. Here $P(C_{\ell})$ is a prior on
the power spectrum, which is chosen to be uniform in the following.
Such mapping may be done very conveniently with Markov Chain Monte
Carlo (MCMC) techniques. Some example CMB applications of MCMC
techniques are described by \citet{knox:2001}, \citet{lewis:2002} and
\citet{eriksen:2006}.

\paragraph{Posterior mapping by Gibbs sampling}

As mentioned above, a direct likelihood evaluation scales as
$\mathcal{O}(N_{\textrm{pix}}^3)$, and the two above algorithms are
therefore limited to low resolution maps. Fortunately, it is possible
to circumvent this problem through a technique called Gibbs sampling
that allows for sampling from $P(C_{\ell}, \mathbf{s} | \mathbf{d})$,
$\mathbf{s}$ being the CMB signal, through the conditional
distributions $P(C_{\ell} | \mathbf{s}, \mathbf{d})$ and $P(\mathbf{s}
| C_{\ell}, \mathbf{d})$.  The conditional nature of the algorithm
avoids inversion of large $\mathbf{S} + \mathbf{N}$ matrices that are
dense in all bases.  It thereby achieves the same goal as the above
MCMC technique with drastically reduced demand on computational
resources \citep{jewell:2004,wandelt:2004,eriksen:2004}.

\paragraph{MASTER}

Approaching the power spectrum estimation problem from a fundamentally
different angle, the pseudo-$C_{\ell}$ methods (e.g., MASTER; Hivon et
al.\ 2002) simply 1) compute the spherical harmonics expansion with
partial sky data; 2) form a raw quadratic estimate of the power
spectrum; 3) correct for noise; and 4) finally decouple the mode
correlations by means of an analytically computable coupling
kernel. Such methods have proved to be very useful tools for CMB
analysis, primarily due to their low computational cost, which again
allows for massive Monte Carlo simulations.

\paragraph{MASTER with cross-spectra}

For multi-channel experiments, a straightforward and very useful
extension to the original MASTER algorithm is simply to include only
cross-correlations between channels in the power spectrum estimate
\citep{hinshaw:2003}.  While an auto-correlation implementation is
quite sensitive to assumptions in the noise model, such assumptions
only affect the error bars in a cross-spectrum and not the mean.

\paragraph{MASTER with internal foreground cleaning}

For multi-frequency experiments with multiple channels per frequency
it is possible to form a set of foreground-cleaned maps using
different channels within each frequency. The noise contributions are
thus still uncorrelated among several pairs of cleaned maps, and the
cross-correlation MASTER implementation may be applied as described
above even to these maps.  Such a method was implemented by
\citep{saha:2006} using the foreground cleaning method of
\citet{tegmark:2003}. We use this method in the present paper for
monitoring residual foregrounds in the spectra computed with
template-cleaned maps, and will refer to the method as ``MASTER with
internal cleaning'' (MASTERint for short; MASTERext refers to
foreground correction by external templates).

\section{Data}
\label{sec:data}

The \emph{WMAP} temperature data \citep{hinshaw:2006} are provided in
the form of ten sky maps observed at five frequencies between 23 and
94 GHz.  These maps are pixelized at HEALPix resolution parameter
$N_{\textrm{side}} = 512$ resulting in about 3 million pixels per
map. In this paper we mainly consider the V-band (61 GHz) and W-band
(94 GHz) channels since these are the ones with the lowest foreground
contamination levels. However, the full data set is used by MASTERint.

We take into account the (assumed circularly symmetric) beam profile
of each channel independently, and we adopt the Kp2 sky cut as our
base mask. This excludes 15.3\% of the sky including all resolved
point sources. However, for the low-resolution analyses we
consider downgraded and/or expanded versions of this mask. The details
of the degradation procedure will be discussed later.

The noise is modeled as uncorrelated, non-uniform and Gaussian with an
RMS given by $\sigma_{0,i} / \sqrt{N_{\textrm{obs}}(p)}$. Here
$\sigma_{0,i}$ is the noise per observation for channel $i$, and
$N_{\textrm{obs}}(p)$ is the number of observations in pixel $p$.
However, this approximation is not adequate for all channels, and in
particular the W-band must be treated with special care because of
significant noise correlations.

We consider two different foreground correction procedures. First, we
simply use the template-corrected maps as provided on the LAMBDA
website, from which a free-free template \citep{finkbeiner:2003}, a
dust template \citep{finkbeiner:1999} and the K-Ka \emph{WMAP} sky map have
been fitted and subtracted. Second, to cross-check these results we
include a MASTER implementation that does not rely on any external
information \citep{saha:2006}. 

Finally, the power spectrum contributions from unresolved point
sources are subtracted band-by-band as a post-processing step
according to the model described by \citet{hinshaw:2006}.

\section{Large-scale analysis}
\label{sec:large_scale}

\begin{figure}[t]

\mbox{\epsfig{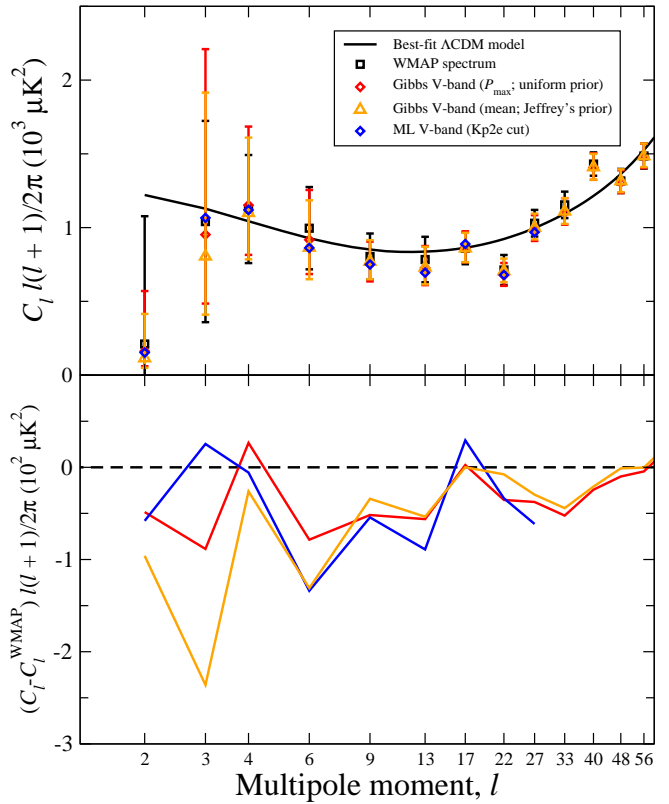}}

\caption{Top panel: Comparison of low-$\ell$ power spectra computed
  from the three-year \emph{WMAP} data using different techniques and
  data combinations. Bottom panel: Differences between the spectra
  computed in this paper and the (appropriately binned) \emph{WMAP}
  spectrum. See text for full details.}
\label{fig:powspec_lowl}
\end{figure}

In this section we present the results from our large-scale analyses,
broadly defined by $\ell \lesssim 50$. The high-resolution results are
shown in Section \ref{sec:small_scale}.

\subsection{Pre-processing and data selection}

Since pixel-space likelihood evaluations scale as
$\mathcal{O}(N^3_{\textrm{pix}})$, we require low-resolution data for
both the maximum-likelihood and MCMC analyses. For these methods, we
therefore consider downgraded maps at $N_{\textrm{side}}=16$, each
having 3072 pixels on the full sky. For Gibbs sampling, angular
resolution does not carry a similar resolution-dependent computational
cost, and the analysis is made with full-resolution data.  MASTER
results are ignored until the next section, since they are sub-optimal
and difficult to compare with other estimates at large angular
scales. However, we note that we have reproduced the MASTER spectra
presented by the WMAP team exactly with MASTERext, and very closely
with MASTERint.

The full-resolution sky maps were downgraded as follows. (See Appendix
\ref{app:lowl_likelihood} for a thorough discussion of this
procedure.) First, the full resolution maps were de-convolved by the
instrumental beam and $N_{\textrm{side}}=512$ pixel window, and then
re-convolved by a Gaussian $9^{\circ}$ FWHM beam and
$N_{\textrm{side}}=16$ pixel window. 
Finally, uniform Gaussian noise with a standard deviation of
$1\,\mu\textrm{K}$ was added to each pixel. This ensures that the map
is properly bandwidth limited at $\ell_{\textrm{max}} = 3\cdot
N_{\textrm{side}} = 48$ and noise dominated beyond $\ell = 40$.

The full-resolution Kp2 mask was downgraded in two different ways. In
the first case, we set excluded pixels to zero and included pixels to
unity. Then we downgrade the mask using the hierarchical structure of
HEALPix, and set each low-resolution pixel to the average of all
sub-pixels. If this average is larger than 0.5, the low-resolution
pixel is accepted. This is the method used in the \emph{WMAP}
likelihood code, and will be denoted by ``Kp2'' in the following. (In
order to compare directly to the \emph{WMAP} results, we first
downgrade to $N_{\textrm{side}}=8$, and then upgrade to
$N_{\textrm{side}}=16$, ensuring identical sky masks.) In the second
case, we smooth the original mask by a Gaussian beam of $9^{\circ}$
FWHM, project the smoothed field onto the low-resolution grid, and
reject all pixels with a value smaller than 0.991, roughly expanding
the original mask by one FWHM in all directions. This mask will be
denoted ``Kp2e'' in the following -- ``e'' for extended -- and
excludes about 7\% more pixels than the Kp2 mask.

We consider three maps in these analyses, namely the de-biased
Internal Linear Combination (ILC) map and the template-corrected V-
and W-band maps individually. Instrumental noise is negligible at
these scales, and co-addition is neither required nor desirable. The
non-zero frequency range is instead used for foreground
monitoring. 

In order to further minimize the impact of foregrounds, we adopt the
approach of the \emph{WMAP} team, and construct a foreground template
as the difference between the raw V-band and the ILC map. This
template is then downgraded in a similar manner as the data maps, but
without noise addition. A corresponding term is added to the
covariance matrix with a large pre-factor in order to remove any
sensitivity to fluctuations with the same spatial pattern as the
template (e.g., Bond et al.\ 1998 and Appendix
\ref{app:lowl_likelihood}).

\subsection{Specification of algorithms}

Three different algorithms are used for the low-$\ell$ analysis,
namely Gibbs sampling, Maximum Likelihood (ML) estimation with
low-resolution data, and MCMC with low-resolution data. In the first
case, two different implementations are used (Commander and MAGIC;
Eriksen et al.\ 2004) using different priors. In both cases, the power
spectra were refined using the Blackwell-Rao estimator
\citep{chu:2005}. For Commander, we use uniform priors, and report the
modes of the posterior distributions as the power spectrum estimates.
This may be directly compared to the \emph{WMAP} spectrum which is
claimed to be a maximum-likelihood estimate at low $\ell$'s. For
MAGIC, we use the Jeffrey's ignorance prior, and report the means as
our power spectrum estimates. In either case, we adopt asymmetric 68\%
confidence regions as our uncertainties.

The ML estimates are established by means of two different non-linear
search algorithms from a commercial library, one quasi-Newton method
and one sequential quadratic programming method.  The free parameters
are the \emph{binned} $C_{\ell}$'s (adopting the binning scheme used
by the \emph{WMAP} team) up to $\ell = 30$, for a total of 9 free
variables.  From $\ell=31$ the spectrum is fixed at the \emph{WMAP}
spectrum values. The covariance matrix is defined as detailed in
Appendix \ref{app:lowl_likelihood}, and consists of a sum of a CMB
signal term, a noise term and a foreground (monopole, dipole and
ILC--V difference map) term.

\begin{figure}[t]

\mbox{\epsfig{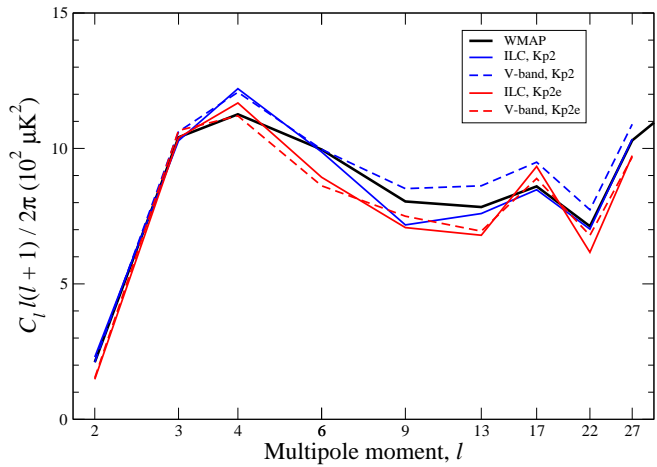}}

\caption{Power spectrum dependence on sky cut and data set. Colored
  lines show the maximum likelihood spectra (computed with a
  pixel-based likelihood) for different sky cuts (blue -- directly
  downgraded Kp2; red -- expanded Kp2) and data sets (dashed -- ILC;
  solid -- V-band). Note that the expanded Kp2 mask spectra lies
  systematically below the directly downgraded Kp2 mask spectra.}
\label{fig:powspec_lowl_cut}
\end{figure}

The MCMC analysis is identical to the ML analysis in terms of the
likelihood evaluation method, binning scheme, multipole range etc.,
but uses a Metropolis-Hastings algorithm to probe the full posterior
distribution. We adopt a uniform prior even in this case. Eight
independent chains are run for 100\,000 steps each, ensuring excellent
convergence properties. One single step costs approximately 1 CPU
second for a 2000 pixel data set.

\subsection{Results}

In Figure \ref{fig:powspec_lowl} we show a comparison of four
different low-$\ell$ power spectrum estimates: The \emph{WMAP}
spectrum (black), the Commander posterior mode uniform prior V-band
spectrum (red), the MAGIC posterior mean Jeffrey's prior V-band
spectrum (orange), and the ML V-band spectrum computed with the Kp2e
mask (blue). In the bottom panel, we show the differences taken
between these spectra and the \emph{WMAP} spectrum. Error bars are
only shown in the top panel: All analyses study the same data set, and
cosmic variance uncertainties can therefore not be used to compare the
relative agreement between the various solutions.

First we note that the individual variations between all four spectra
are about 50-100$\,\mu\textrm{K}^2$, corresponding to about 5-10\% of
the absolute spectrum amplitude. This is much less than the cosmic
variance, and the analysis uncertainty per mode is thus not a dominant
effect. However, it is troublesome that this difference appears to be
systematic: All three colored spectra lie below the \emph{WMAP}
spectrum up to $\ell \approx 30-40$. Since the cumulative effect over
an entire multipole decade of even a small bias may potentially affect
cosmological parameters, it is important to determine the origin of
this discrepancy.

Foreground uncertainties are always a concern at large angular scales
for CMB experiments and a most natural first candidate to
consider. First, we note that the four spectra shown in Figure
\ref{fig:powspec_lowl} are computed from slightly different data sets:
The \emph{WMAP} spectrum is based on the ILC map with a directly
downgraded (i.e., not expanded) Kp2 mask for $\ell \le 12$ and on the
template-corrected V- and W-bands with the high-resolution Kp2 cut at
$\ell > 12$; the Gibbs spectra are based on the template-corrected
V-bands and the full-resolution Kp2 cut at all scales; the ML estimate
is computed from the degraded template-corrected V-band, cut with the
expanded Kp2 mask.



To assess the impact of residual foregrounds in the low-$\ell$
power spectra, we show in Figure \ref{fig:powspec_lowl_cut} the ML
power spectra computed from the ILC and template-corrected V-band
maps, with both the original Kp2 and extended Kp2e sky cuts. A
noticable trend is clearly visible in this range, in that there is a
significant loss of power between the Kp2 and Kp2e cut for both the
ILC and V-band maps. This implies that there is considerable
additional power in the 7\% near-galactic pixels in the Kp2 sky cut
relative to the Kp2e cut for both maps, and both maps are most likely
contaminated at some level near the mask boundaries. On the other
hand, after expanding the mask the two spectra are quite similar,
possibly indicating that one is fairly safe after expanding the mask,
and that the high latitude residuals are small compared to the CMB
fluctuations.

\begin{figure*}[t]

\mbox{\epsfig{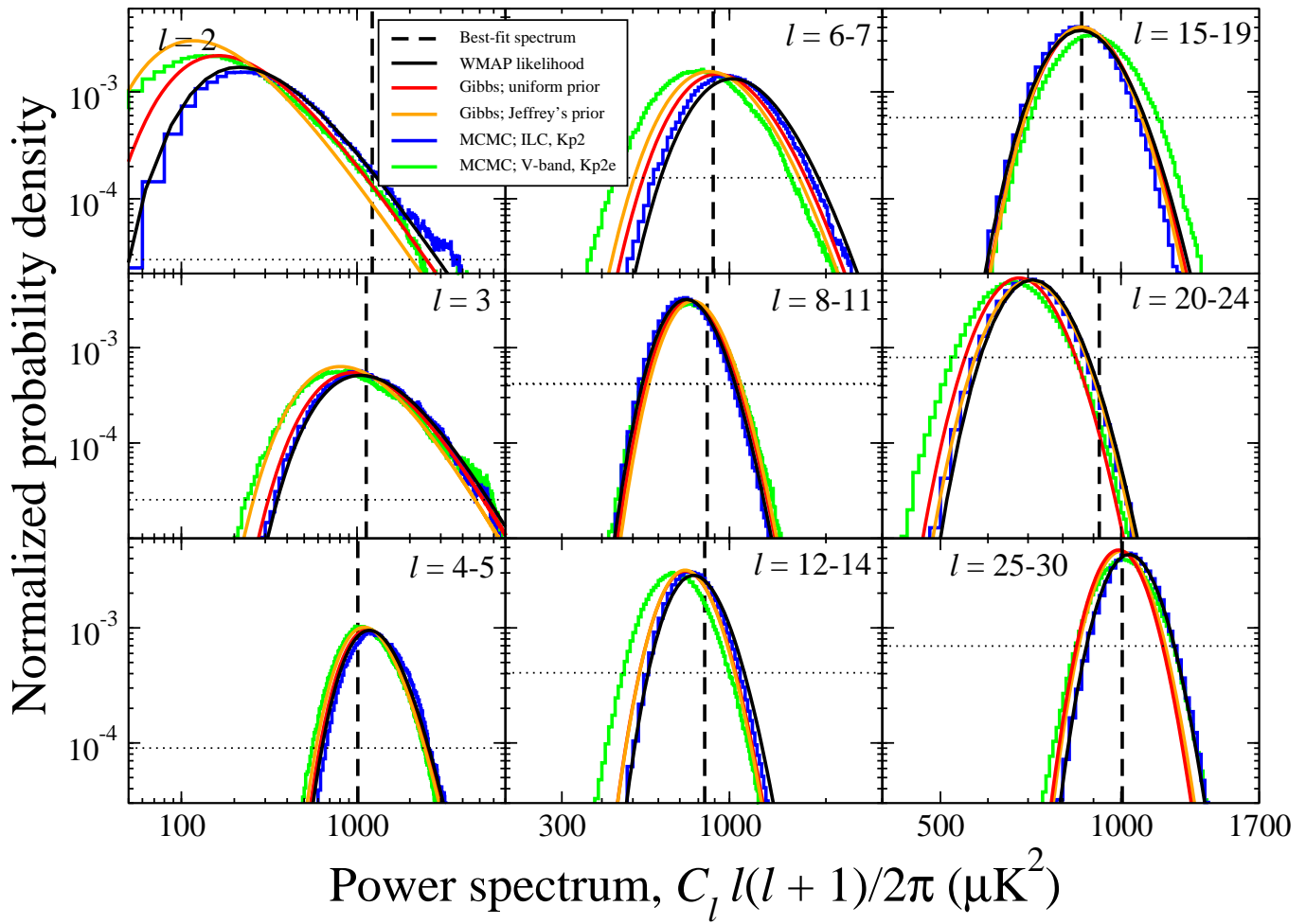}}

\caption{Posterior distributions (and the \emph{WMAP} likelihood)
  computed by different methods and data sets. The colored histograms
  show the results from MCMC runs with low-$\ell$ likelihood computed
  in pixel space for different masks, and the black curve shows the
  Blackwell-Rao estimator computed with V-band/Kp2 Gibbs samples. The
  vertical dashed lines show the angular power spectrum for the
  $\Lambda$CDM model that best fits the \emph{WMAP} data
  \citep{hinshaw:2006}. The horizontal dashed lines shows the 95\%
  confidence region relative to the Gibbs-based estimate. Note that
  while the MCMC shows marginalized distributions, the other curves
  indicate slices through the multi-dimensional distributions with
  other $C_{\ell}$'s fixed at the best-fit spectrum value.}
\label{fig:likelihoods}
\end{figure*}

While the mask explanation can account for some of the discrepancies
seen in Figure \ref{fig:powspec_lowl}, it only does so at $\ell \le
12$ where a pixel-based likelihood evaluation is used by the
\emph{WMAP} team. For $\ell > 12$ a MASTER-based likelihood is used
which is based on the full-resolution template-corrected data. The
small discrepancies seen in the range between $12 < \ell \lesssim 30$
is therefore due to differences in statistical treatment, and not data
selection. 

A main question is therefore the following: For what $\ell$-range
does the MASTER-based likelihood approximation provide an acceptable
fit to the exact likelihood? From Figure \ref{fig:powspec_lowl} it
seems clear that $\ell_{\textrm{exact}} \le 12$ is not sufficient,
while $\ell_{\textrm{exact}} \le 50$ is quite likely more than
enough. It is difficult to answer this question based on a power
spectrum plot alone, and we will therefore return to this question in
Section \ref{sec:parameters} where we estimate cosmological parameters
with various likelihoods. 

In order to further illuminate the above issues, we show in Figure
\ref{fig:likelihoods} a set of different likelihood (and posterior)
distributions computed from the \emph{WMAP} data. The vertical dashed
lines show the binned best-fit $\Lambda$CDM spectrum and the curves
show the \emph{WMAP} likelihoods (black), Gibbs + Blackwell-Rao
posterior distributions (red/orange) and MCMC posterior distributions
(blue/green), respectively. Two sigma confidence regions relative to
the Gibbs distributions with uniform priors (red curves) are indicated
by horizontal dotted lines.

\begin{figure}[t]

\mbox{\epsfig{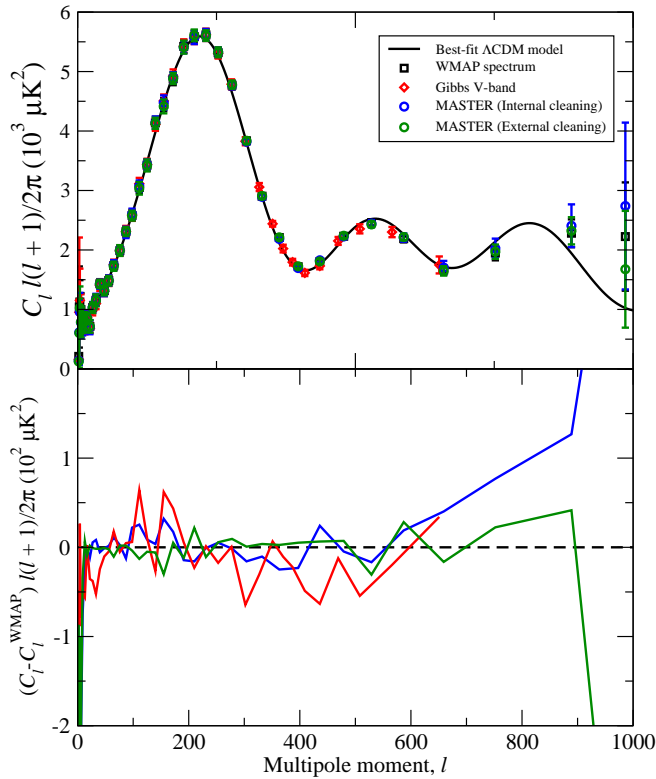}}

\caption{Top panel: Comparison of power spectra computed from the
  three-year \emph{WMAP} data using different techniques and data
  combinations. Bottom panel: Differences between the spectra computed
  in this paper and the (appropriately binned) \emph{WMAP}
  spectrum.}
\label{fig:powspec}
\end{figure}

Many interesting points may be seen in this figure:
\begin{enumerate}
\item Comparing the V-band+Kp2e and ILC+Kp2 MCMC runs (green versus
  blue) we see that the former is shifted systematically to lower
  values of $C_{\ell}$.  This shift is most likely due to a
  combination of extra fluctuation power coming from the region
  excluded by Kp2e but included by Kp2, and additional sampling
  variance from the extended sky cut.
\item The agreement between the ILC+Kp2 MCMC and \emph{WMAP} likelihood
  distributions (blue versus black) is generally quite reasonable, but
  small shifts may be seen in a number of cases, especially for $6 \le
  \ell \le 12$. This is likely due to different likelihood
  regularization (see Appendix \ref{app:lowl_likelihood}): We use a
  properly bandwidth limited map within the Nyquist frequency
  $\ell_{\textrm{N}} = 3N_{\textrm{side}}$ but add a small noise term
  to regularize the covariance matrix; the \emph{WMAP} code increases
  $\ell_{\textrm{max}}$ to $4N_{\textrm{side}}$, well beyond the
  Nyquist frequency, to make the covariance matrix non-singular.
\item The V-band+Kp2e MCMC distributions are generally wider than the
  ILC+Kp2 distributions (green versus blue) because of a more
  conservative mask. This is avoided with the full-resolution Gibbs
  analysis (red curve), since it uses unsmoothed data and thereby
  does not need to expand the mask. Consequently, the Gibbs
  distributions are about as wide as the ILC+Kp2 distributions. 
\item The Jeffrey's ignorance prior puts more probability on small
  values of $C_{\ell}$, but is mostly relevant at very low $\ell$'s
  (orange versus red). Because of the very large cosmic variance of
  these multipoles, the effect of the Jeffrey's prior on
  cosmological parameters is small. 
\item The quadrupole posterior maximum with Jeffrey's and uniform
  priors are respectively $105\mu\textrm{K}^2$ and
  $160\mu\textrm{K}^2$. Thus, even with Jeffrey's prior the
  \emph{WMAP} quadrupole amplitude is completely unremarkable relative
  to the best-fit $\Lambda$CDM value, and always well inside the 95\%
  confidence region (vertical dashed versus horizontal dotted
  lines). The most extreme outlier is that of $20 \le \ell \le 24$,
  whose value is low at the $99.4\%$ level.
\end{enumerate}


\section{Intermediate- and small-scale analysis}
\label{sec:small_scale}

We now consider the intermediate- and small-scale parts of the power
spectrum.

\subsection{Algorithms and data selection}

Since we study full-resolution data in this section, the direct
likelihood evaluation techniques are no longer available to
us. However, on these scales the MASTER algorithms are applicable, and
we once again have multiple methods available for cross-checking
purposes. In total four different codes are used, namely Commander
(Gibbs sampling; uniform prior), MAGIC (Gibbs sampling; Jeffrey's
prior), MASTERext and MASTERint.

For the Gibbs analyses we consider only the V-band data; the Q-band is
considered to be too foreground contaminated for reliable analysis,
and the W-band exhibits strong correlated noise that significantly
biases any auto-correlation method \citep{hinshaw:2006}. This has been
verified in our analyses, and we exclude these bands entirely from the
Gibbs analyses. Next, we analyzed the data with both Commander and
MAGIC, and obtained identical results (up to convergence) on small and
intermediate scales where the prior is less important. Therefore we
show only one set of Gibbs results in the following.

In the case of the MASTER analyses, we include cross-spectra only in
the final power spectra, and are thus quite safe with respect to
correlated noise. For this reason the W-band is included in these
cases.  Further, for the internal cleaning analysis two different data
combinations were considered, namely including either all five bands
or only the cleanest Q-, V- and W-bands. The difference between these
two spectra was found to be very small. We once again present only one
(the five-band) solution here, and note that this solution is not
strongly dependent on the low-frequency bands.

The foreground mask is always the full-resolution Kp2 that excludes
resolved point sources. Contributions to the power spectrum from
unresolved point sources are subtracted band-by-band according to the
model described by \citet{hinshaw:2006}.

\subsection{Results}

A summary of our main intermediate- and small-scale power spectra are
shown in Figure \ref{fig:powspec}. Four spectra are plotted here: The
\emph{WMAP} spectrum, the Gibbs V-band spectrum, the V+W MASTERext
cross-spectrum, and the five-band MASTERint cross-spectrum.

In the top panel we see that the overall agreement is excellent
considering the very different approaches taken by the different
methods. It is thus very unlikely that algorithmic issues play a
dominant role in the determination of cosmological parameters as
presented by \citet{spergel:2006} for the small-scale temperature
anisotropy results.

In the bottom panel we plot the difference between each of the colored
spectra and the \emph{WMAP} spectrum in order to look for systematic
relative trends. Again, we see that the overall agreement is very
good, and the only possible anomaly is a slight power deficit for the
V-band Gibbs spectrum at $\ell \approx 300$--600. This will be studied
further shortly. 

The extreme high-$\ell$ discrepancies seen in two MASTER spectra are
due to noise weighting differences. For the MASTERext analysis, we
obtained the individual cross-spectra from the WMAP team and verified
that our spectra are identical to those of the WMAP team. However,
while the \emph{WMAP} spectrum uses an elaborate and nearly
maximum-likelihood weighting scheme for co-adding these cross-spectra
into one optimal spectrum, we use either simple inverse-variance
weighting (for MASTERext) or flat weighting (for
MASTERint). Differences in the very low signal-to-noise regime are
thus not unexpected.

The excellent agreement seen in Figure \ref{fig:powspec} is perhaps
most remarkable in the case of the MASTERint implementation: While all
other spectra more or less correspond to different statistical
treatment of the same basic data set, this particular solution takes a
drastically different approach, and uses only \emph{WMAP} data by
themselves to correct for foregrounds. Thus, it provides an important
and impressive cross-check on the \emph{WMAP} foreground cleaning
approach.

\begin{figure}[t]

\mbox{\epsfig{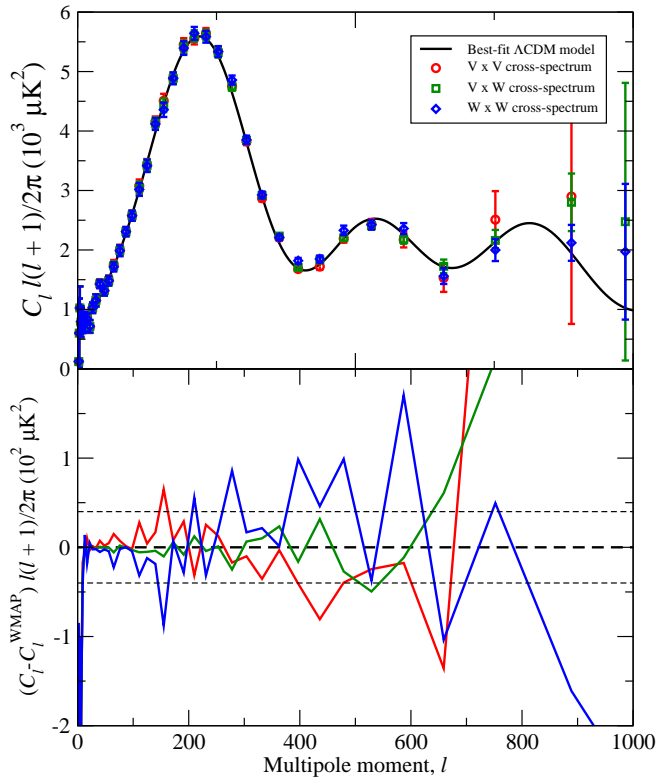}}

\caption{Power spectrum versus frequency. All spectra are computed
  from the template-corrected maps with the basic cross-MASTER
  algorithm. Only the frequency contents vary, as each spectrum only
  contains cross-information of V$\times$V, V$\times$W and W$\times$W
  respectively.}
\label{fig:powspec_freq}
\end{figure}

However, returning to the power deficit issue seen in the V-band
spectrum, we plot in Figure \ref{fig:powspec_freq} the MASTERext
spectra for individual frequency combinations, including V$\times$V,
V$\times$W, or W$\times$W spectra, respectively. In this plot we see
that the MASTER V-band spectrum agrees very well with the Gibbs V-band
spectrum, and lies systematically below the averaged \emph{WMAP}
spectrum between $\ell \approx 300$ and 600. Furthermore, we see that
the W-band spectrum lies systematically above the \emph{WMAP} spectrum
in the same range, with a relative bias of up to
$80\,\mu\textrm{K}^2$. The V$\times$W cross-spectrum lies in between,
being more or less an average of the two.

To quantify the significance of this difference, we analyzed the set of
2500 Monte Carlo simulations that was used for the MASTERext
analysis. For each simulation we first computed the difference between
the W- and V-band spectra, and then the mean (inverse noise variance
weighted) difference within some $\ell$-range. Comparing the observed
\emph{WMAP} data to these simulations, we found that the discrepancy
is significant at slightly more than $3\sigma$ for both $250 \le \ell
\le 600$ and $400 \le \ell \le 500$, corresponding to the largest
region of asymmetry and to the most discrepant region,
respectively. Presumably it is possible to boost the significance by
tuning the region further, but on the other hand, the known $\sim1\%$
beam amplitude uncertainty that is taken into account by the WMAP
likelihood code will reduce the signficance by a few tenths of a
sigma. Independent of such minor effects, it seems that the
probability of this being a statistical fluke is rather small.

After the publication of the present paper, a follow-up study by
\citet{huffenberger:2006} revisited the unresolved point source
analysis performed by the \emph{WMAP} team. The main result from that
work was a best-fit unresolved point source spectrum amplitude of $A =
0.011 \,\mu\textrm{K}^2 \textrm{sr}$ (relative to 41 GHz), which is
significantly lower than the value of $A = 0.017 \,\mu\textrm{K}^2
\textrm{sr}$ initially reported \citep{hinshaw:2006} and used in the
present paper. Thus, a relatively larger contribution is subtracted
from the V-band than from the W-band spectrum, and in effect, the
V-band spectrum has been over-corrected. After taking into account
this new amplitude, the discrepancy between the two band spectra was
found to be significant at $\sim2\sigma$ using the same test as
above.


While the quoted point source correction can account for a
substantial amount of the observed discrepancy, there is still a small
difference present, and this could possibly indicate further
systematics. In this respect, it could be worth considering unmodeled
beam asymmetries. As a preliminary step in the three-year analysis,
\citet{hinshaw:2006} considered the impact of such asymmetries on the
individual cross-spectra. After an extensive analysis, they concluded
that their magnitude is less than 1\% at $\ell \le 1000$, and
therefore sufficiently small to neglect in further analyses. However,
at the relevant scales, the absolute amplitude of the temperature
power spectrum is about $2000\,\mu\textrm{K}^2$, and a beam asymmetry
bias of 1\% therefore corresponds to an expected discrepancy of
$20\,\mu\textrm{K}^2$. This corresponds roughly to $1\sigma$ of the V-
versus W-band difference, and, if determined appropriate, its
correction could bring the overall difference well within the
statistical errors.
Fortunately, this issue will be clearer with additional years of
\emph{WMAP} observations.

\section{Cosmological parameters}
\label{sec:parameters}

In the previous two sections we considered the temperature angular
power spectrum as observed by \emph{WMAP}, and our main conclusion
from these analyses is that the \emph{WMAP} spectrum is reasonable at
all angular scales. However, there are small but clearly noticeable
biases at both large and small scales. In this section, we seek
to quantify the impact of this bias in terms of the cosmological
parameters for a minimal six-parameter $\Lambda$CDM model. The
combined effect of both the low- and high-$\ell$ discrepancies are
studied by \citet{huffenberger:2006}, and the effect on extended
cosmological models (e.g., models including massive neutrinos and
running of the spectral index) are considered by
\citet{kristiansen:2006}.

\begin{deluxetable*}{cccccc}
\tablewidth{0pt} 
\tabletypesize{\small} 
\tablecaption{Cosmological parameters\label{tab:parameters}}
\tablecolumns{6}
\tablehead{
& & \emph{WMAP} + BR & \emph{WMAP} + BR & \emph{WMAP} + Pixel & Shift \\
Parameter & \emph{WMAP} & ($\ell_{\textrm{max}} = 12$) &
($\ell_{\textrm{max}} = 30$) & ($\ell_{\textrm{max}} = 30$) & (in $\sigma_{\textrm{WMAP}}$)  
}

\startdata

\cutinhead{\emph{WMAP} data only}
 $\Omega_{\textrm{b}}\,h^2$    & $0.0222 \pm 0.0007$ &
 $0.0222\pm0.0007$ & $0.0224\pm0.0007$ & $0.0224\pm0.0007$ &  -0.3 \\ 
 $\Omega_{\textrm{m}}$    & $0.241 \pm 0.036$ & $0.241\pm0.0037$ & $0.230\pm 0.033$
 & $0.234\pm0.035$& \phm{-}0.3 \\ 
 $\log(10^{10}A_{\textrm{s}})$    & $3.019 \pm 0.067$ & $3.017\pm0.068$&
 $3.014\pm 0.068$ & $3.013\pm0.068$& \phm{-}0.1 \\ 
 $h$    & $0.731 \pm 0.033$ & $0.731\pm0.032$ & $0.743\pm 0.033$
 & $0.739\pm0.033$& -0.4 \\ 
 $n_{\textrm{s}}$    & $0.954 \pm 0.016$ & $0.955\pm0.016$& $0.961\pm 0.016$
 & $0.960\pm0.017$& -0.4  \\ 
 $\tau$    & $0.090 \pm 0.030$ & $0.088\pm0.030$& $0.090\pm 0.030$
 & $0.088\pm0.030$& \phm{-}0.0  \\

\cutinhead{\emph{WMAP} + Acbar + BOOMERanG}
 $\Omega_{\textrm{b}}\,h^2$    & $0.0225 \pm 0.0007$ &
 $0.0225\pm0.0007$ & $0.0227\pm0.0007$ & $0.0227\pm0.0007$ & -0.3  \\ 
 $\Omega_{\textrm{m}}$    & $0.239 \pm 0.031$ & $0.240\pm0.032$ & $0.229\pm 0.030$
 & $0.233\pm0.031$ & \phm{-}0.3  \\ 
 $\log(10^{10}A_{\textrm{s}})$    & $3.030 \pm 0.064$ &
 $3.028\pm0.065$ &  $3.025\pm 0.066$ & $3.023\pm0.0065$ & \phm{-}0.1 \\ 
 $h$    & $0.737 \pm 0.029$ & $0.736\pm0.031$ & $0.749\pm 0.031$
 & $0.744\pm0.031$ & -0.4  \\ 
 $n_{\textrm{s}}$    & $0.958 \pm 0.016$ & $0.959\pm0.016$ & $0.965\pm 0.016$
 & $0.964\pm0.016$ & -0.4  \\ 
 $\tau$    & $0.091 \pm 0.030$ & $0.090\pm0.030$  & $0.091\pm 0.030$
 & $0.089\pm0.030$ & \phm{-}\phm{-}0.0
\enddata

\tablecomments{Comparison of marginalized parameter results obtained
  from the \emph{WMAP} likelihood (second column), the \emph{WMAP} +
  Blackwell-Rao hybrid (third and fourth columns), and from the
  \emph{WMAP} + $N_{\textrm{side}}=16$ pixel-based likelihood (fifth
  column). The latter three were based on the template-corrected
  V-band data at low $\ell$'s. The relative shift between the
  \emph{WMAP} and the $\ell_{\textrm{max}}=30$ BR hybrid (in units of
  $\sigma_{\textrm{WMAP}}$) is shown in the rightmost column.}

\end{deluxetable*}

We perform four sets of similar analyses, all primarily based on
the \emph{WMAP} likelihood code. First, we adopt the \emph{WMAP}
likelihood as provided. Second, we replace the low-$\ell$ likelihood
(both the pixel-based estimator and the low-$\ell$ MASTER estimator)
with a Blackwell-Rao (BR) Gibbs-based estimator for $\ell \le 30$
\citep{chu:2005}. Third, we do the same for $\ell \le 12$
alone. Finally, we use the $N_{\textrm{side}} = 16$,
$\ell_{\textrm{max}} = 30$ pixel-based likelihood for the V-band and
Kp2e cut described earlier.  For all cases, we analyze two data
combinations; the \emph{WMAP} data alone, and the combination of
\emph{WMAP}, Acbar \citep{kuo:2004} and BOOMERanG
\citep{montroy:2005,piacentini:2005,jones:2005}. Marginalization of SZ
was not included.

\begin{figure}[t]

\mbox{\epsfig{figure=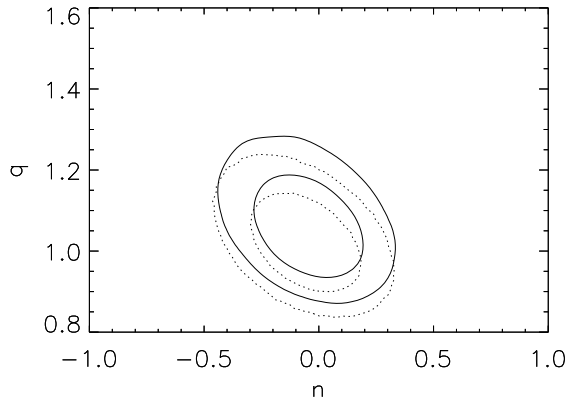,width=\linewidth,clip=}}

\caption{Probability contours for a simple amplitude/tilt model
  ($C_{\ell}(q,n) = q\, C_{\ell}^{\textrm{fid}}
  \left(\ell/10\right)^n$, $C_{\ell}^{\textrm{fid}}$ being the
  best-fit $\Lambda$CMB model) fitted to the \emph{WMAP} likelihood (solid
  lines) and to the Gibbs sampled posterior (dotted lines).  Contours
  indicate 68 and 95\% confidence regions. Note the shift to higher
  amplitudes for the \emph{WMAP} results, and also that the Gibbs results are more
  closely centered on $(q,n) = (1,0)$.  }
\label{fig:qn_comp}
\end{figure}

Note that we use the \emph{WMAP} likelihood at high $\ell$'s in all
cases in order to highlight the effects of the low-$\ell$ likelihood
bias.  While using a Gibbs sampling based estimator even at high
$\ell$'s could potentially have beneficial effects in terms of
uncertainties, it might confuse the low-$\ell$ bias analysis by
introducing other differences.

The cosmological parameters corresponding to these likelihoods were
established through standard Markov Chain Monte Carlo analysis. In
keeping with our philosophy of cross-verification, two independent
codes were used for a few cases. In the first case, CosmoMC
\citep{lewis:2002} was downloaded and appropriately modified, and in
the second, a stand-alone code was written from scratch in C++. The
two codes returned identical distributions, and as usual we show only
one set of results in the following. We also performed similar
analyses using only temperature-data, imposing a Gaussian prior on the
optical depth of $\tau = 0.10\pm0.03$ to simulate the effect of
polarization data, but without relying on the accuracy of these. As
expected, we then obtained very similar results to those reported
here.

Before reporting the physical parameters, we consider a very simple
phenomenological model in order to gain intuition on what to
expect. Specifically, we fit a model with a free amplitude and tilt to
both the \emph{WMAP} likelihood and the \emph{WMAP}+BR
$\ell_{\textrm{max}}=30$ hybrid,
\begin{equation}
C_{\ell}(q,n) = q\, C_{\ell}^{\textrm{fid}} \left(\frac{\ell}{10}\right)^n.
\end{equation}
Here $C_{\ell}^{\textrm{fid}}$ is a fiducial model (chosen to be the
best-fit $\Lambda$CDM power law model), $q$ is a relative amplitude,
and $n$ is a tilt parameter. Since we expect the fiducial model to be
a good fit to the data, we anticipate values of $(q,n)$ around
$(1,0)$. In this exercise we include only data between $2 \le \ell \le
20$. The results from these computations are shown in Figure
\ref{fig:qn_comp}, where contours indicate 68 and 95\% confidence
regions. \emph{WMAP} results are indicated by solid lines and
Blackwell-Rao results by dotted lines.

\begin{figure*}[t]

\mbox{\epsfig{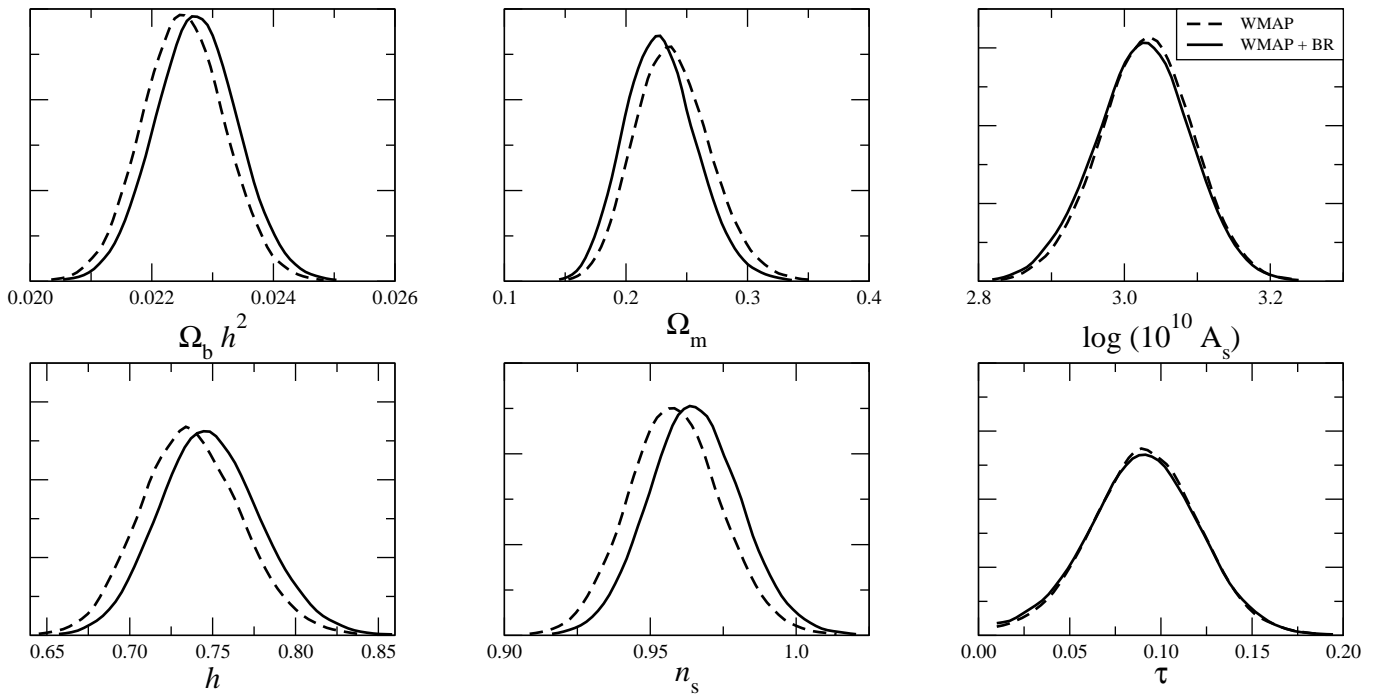}}

\caption{Comparison of marginal probability distributions as computed
  with the \emph{WMAP} likelihood (dashed) and with the \emph{WMAP} +
  Blackwell-Rao hybrid (solid) at $2 \le \ell \le 30$. The
  results are based on the WMAP, Acbar and BOOMERanG data, as
  described in the text.}
\label{fig:parameters}
\end{figure*}

By far the most striking feature is a $\sim0.5\sigma$ shift in
amplitude towards lower values for our revised posterior, consistent
with the low-$\ell$ power spectrum results shown earlier. However, it
is interesting to notice that the Blackwell-Rao contours are more
centered on $(q,n) = (1,0)$ than the \emph{WMAP} contours: This
indicates that there is less tension between the low-$\ell$ and
high-$\ell$ parts of the spectrum when using our approach instead of
the \emph{WMAP} approach.

For the physical parameter estimation, we adopted a standard
six-parameter $\Lambda$CDM model with $\{\Omega_{\textrm{b}}h^2,
\Omega_{\textrm{c}}h^2, \theta, \tau, n_{\textrm{s}}, \log (10^{10}
A_{\textrm{s}})\}$ as our free parameters. The priors were chosen to
be uniform for all parameters.

The results from these computations are summarized in Table
\ref{tab:parameters} and Figure \ref{fig:parameters}, and a number of
interesting points may be seen here. First, comparing the \emph{WMAP}
likelihood to the $\ell \le 12$ \emph{WMAP} + BR hybrid (second and
third columns of Table \ref{tab:parameters}), we see that the small
change due to mask expansion at $\ell \le 12$ seen in Figure
\ref{fig:powspec_lowl_cut} has no impact in terms of cosmological
parameters. In effect, the cosmic variance on these very large scales
hides such problems.

However, comparing the \emph{WMAP} likelihood to the $\ell \le 30$
\emph{WMAP} + BR hybrid (second and fourth columns), a different
picture is seen. In this case, we find shifts of up to $0.4\sigma$ for
both $h$ and $n_{\textrm{s}}$ (right column). Thus, the MASTER-based
approximation used in this regime appears to be inadequate for the
precision required by the \emph{WMAP} data. That the differences are
indeed due to statistical treatment is confirmed by the results for
the low-$\ell$ pixel-based hybrid, which extends the original
\emph{WMAP} analysis simply by using direct evaluation up to $\ell=30$
instead of $\ell=12$. 

We also performed a second analysis with the \emph{WMAP} + BR hybrid,
this time making the transition at $\ell=50$. Fortunately, the results
from these computations were essentially identical to those for
$\ell=30$, and this implies that a pixel-based likelihood for
$N_{\textrm{side}}=16$ may still be used for the precision of the
\emph{WMAP} data. However, the computational expense of this approach
is already pushing close to what is reasonable, and currently
dominates the MCMC cost. This illustrates very well the advantages of
the Gibbs sampling approach, in which identical results are obtained
at only a fraction of the computational cost.

In Figure \ref{fig:parameters}, we compare the results for the
\emph{WMAP} likelihood and \emph{WMAP} + BR hybrid likelihood from the
combined \emph{WMAP}, Acbar and BOOMERanG analysis in terms of
marginalized probability distributions. With a few exceptions, the
low-$\ell$ bias results in a shift of a few tenths of a sigma. Perhaps
most interesting is the effect on the much debated $n_{\textrm{s}}$:
This distribution is shifted towards higher values by the new
low-$\ell$ likelihood, and the evidence for $n_{\textrm{s}} \ne 1$ is
thus reduced.




Finally, we note that with Planck's reach to higher $\ell$'s, the $2
\le \ell \lesssim 30$ data will not be critically important for
constraining $n_{\textrm{s}}$; removing $\ell \le 30$ will increase
$\sigma(n_{\textrm{s}})$ by less than 10\%.  Planck's constraints on
$n_{\textrm{s}}$ will thus not be dependent on low-$\ell$ polarization
measurements, and therefore less sensitive to residual foregrounds.


To summarize, we find that the improvements we make to the low-$\ell$
likelihood have a small but noticeable impact on the cosmological
parameters reported by \citet{spergel:2006}. In particular, the
evidence for $n_{\textrm{s}} \ne 1$ is weakened by $0.4\sigma$, which
is significant given the initial marginal $\sim2.5-3\sigma$
detection. For full details on results correcting for both this
low-$\ell$ statistical issue and the high-$\ell$ point source issue,
we refer to \citet{huffenberger:2006} and \citet{kristiansen:2006}.

\section{Conclusions}
\label{sec:conclusions}

In this paper we have made an extensive re-analysis of the \emph{WMAP}
temperature anisotropy data. We have re-computed the power spectrum
using six different codes, and the cosmological parameters with two
different codes. Five different groups have contributed with separate
computations. In total, the amount of cross-checks provided by this
effort leads us to conclude that most important analysis issues
concerning the three-year temperature power spectrum are well
understood.

Our main conclusion from this work is that we confirm the \emph{WMAP}
temperature power spectrum on most scales. However, subtle
discrepancies are found both at large and small scales, and these can
be summarized as follows:
\begin{enumerate}
\item There is a small but significant bias at large angular scales
  (5--10\% at $\ell \le 30$) in the \emph{WMAP} power spectrum and
  published likelihood code. This is primarily due to statistical
  issues in the form of an inadequate likelihood approximation, and
  secondarily due to residual foregrounds.
\item There is a systematic bias between the V- and W-bands. This
  has recently been identified by \citet{huffenberger:2006} to be at
  least partially caused by over-correction of unresolved point
  sources. However, even after correcting for this, a small
  discrepancy is present, and this could possibly indicate further
  systematics. A possible candidate worth considering could be
  uncorrected beam asymmetries.
\item The low-$\ell$ likelihood bias has a noticable effect of some
  parameters. Most interestingly, it increases $n_{\textrm{s}}$ by
  $0.4\sigma$, reducing the nominal detection of $n_{\textrm{s}} \ne
  1$ from $\sim2.7$ to $\sim2.3\sigma$. Further, as reported by
  \citet{huffenberger:2006}, correcting for the point source
  over-correction increases $n_{\textrm{s}}$ by an additional
  $\sim0.3\sigma$, to a final significance of $n_{\textrm{s}} \ne 1$
  of $\sim2.0$.
\end{enumerate}

In addition to these cosmologically important results, we make a few
algorithmic recommendations based on the work presented here regarding
the \emph{WMAP} analysis. First and foremost, an exact likelihood
evaluation should be used at least up to $\ell \le 30$. Either direct
evaluation or Gibbs sampling can be used for these scales, but of
course, the negligible computational cost of the latter, when
incorporated into an MCMC sampler, makes it the preferred
choice. Second, we advocate using the more conservative Kp2e cut at
very low l as a hedge against possible foreground contamination just outside the
 directly downgraded Kp2 mask. However, this has a negligible effect in terms of
cosmological parameters.

The products from the Gibbs sampling analysis (with a uniform prior)
are made publically available\footnote{Available from
http://www.astro.uio.no/$\sim$hke/wmap3} in the form of three data
sets. First, the most relevant for the applications presented in this
paper is the collection of $4\times 1000$ sky signal spectrum
samples. These form the basis of the Blackwell-Rao estimator, and may
be used for parameter estimation as demonstrated in this paper. A
corresponding Fortran 90 module is also provided. Second, a set of
4000 ensemble averaged spectrum samples are presented. These may be
used for visualization purposes of the power spectrum. Third, a set of
sky map samples are provided for $\ell \le 50$. These may be used for
analyses that require phase information, for instance studies of
non-Gaussianity.



Thus, we conclude that although fast methods such as MASTER are very
useful in the early stages of analyzing a new experiment, when fast
turn around times are essential, the extensions to Gibbs sampling (as
described in the Appendices) have now rendered an exact method quite
tractable with currently available computing resources. Based upon our
experience with these methods, we believe that Gibbs sampling can and
will play a significant role in the future analysis of Planck data.

\begin{acknowledgements}
  We thank Eiichiro Komatsu and Gary Hinshaw for useful discussions
  and comments, and for their thorough work on reproducing the results
  presented here. We acknowledge use of the HEALPix software
  \citep{gorski:2005} and analysis package for deriving the results in
  this paper. We acknowledge use of the Legacy Archive for Microwave
  Background Data Analysis (LAMBDA). This work was partially performed
  at the Jet Propulsion Laboratory, California Institute of
  Technology, under a contract with the National Aeronautics and Space
  Administration. HKE acknowledges financial support from the Research
  Council of Norway.  BDW acknowledges support through NSF grant
  AST-0507676 and NASA JPL subcontract 1236748. RS, PJ and TS
  acknowledge the use the IUCAA HPC facility for computations. TS
  thanks JPL for supporting the visit that helped initiate the
  collaboration.
\end{acknowledgements}

\appendix


\section{A. Low-$\ell$ likelihood evaluation}
\label{app:lowl_likelihood}

Low-$\ell$ likelihood evaluation from high-resolution data has
received considerable attention in the last few years
\citep{efstathiou:2004, slosar:2004, hinshaw:2006}. The reason is
simply that while the currently popular pseudo-spectrum power spectrum
estimators \citep{hivon:2002} work very well on intermediate and small
angular scales, they are clearly sub-optimal on large scales.

The log-likelihood for a pixelized Gaussian random field $\mathbf{d}$,
with vanishing mean and covariance matrix $\mathbf{C}$, is given by
\begin{equation}
-2\log \mathcal{L} = \mathbf{d}^{t} \mathbf{C}^{-1} \mathbf{d} + \log
|\mathbf{C}|,
\label{eq:lnl}
\end{equation}
up to a normalization constant. Evaluation of this expression requires
inversion of the pixel-pixel covariance matrix, and therefore scales
as $\mathcal{O}(N_{\textrm{pix}}^3)$. Consequently, direct likelihood
analysis of mega-pixel maps is not currently feasible.

The covariance matrix consists of a series of terms depending on the
data model, but at the very least one needs a signal term
$\mathbf{S}$, which for an assumed isotropic CMB is given by the
angular power spectrum $C_{\ell}$,
\begin{equation}
S_{ij} = \frac{1}{4\pi} \sum_{\ell=2}^{\ell_{\textrm{max}}} (2\ell+1)
C_{\ell} b_{\ell}^2 P_{\ell}(\cos \theta_{ij}).
\label{eq:signal_mat}
\end{equation}
Here $b_{\ell}$ is the Legendre transform of the (circularly
symmetric) experimental beam, $P_{\ell}(x)$ are the Legendre
polynomials, $\theta_{ij}$ is the angle between pixels $i$ and $j$,
and $\ell_{\textrm{max}}$ is a sufficiently large multipole. The exact
definition of ``sufficiently large'' will be made explicit shortly.

Next, in most real-world applications there is also a term describing
instrumental noise, $\mathbf{N}$. Often, this is modeled as
independent between pixels (white), and given by an overall noise
level per observation $\sigma_0$ and a total number of observations
per pixel $N_{\textrm{obs}}(i)$, such that $N_{ij} = \sigma_0 /
\sqrt{N_{\textrm{obs}}(i)} \delta_{ij}$. For current CMB experiments,
such as \emph{WMAP}, this is strongly sub-dominant to the signal term on
large angular scales, and may in principle be omitted without
significant loss of accuracy. However, this is not entirely true,
since it can (and should) play an important role in regularizing the
covariance matrix.

Finally, it is useful to include a number of template terms over which
amplitudes one wishes to marginalize \citep{bond:1998,
slosar:2004}. Consider for instance a given foreground template
$\mathbf{f}$ whose spatial structure is known, but overall amplitude
is unknown. Then the corresponding covariance matrix is given by the
outer product $\mathbf{F} = \mathbf{f}\,\mathbf{f}^{t}$, and by
assigning a sufficiently large uncertainty to this structure the
corresponding mode is projected out from the data. In total, the
covariance matrix may be written as
\begin{equation}
\mathbf{C} = \mathbf{S} + \mathbf{N} + \sum_{i} \lambda_i \mathbf{F}_i,
\end{equation}
where $\lambda_i$ are numerically sufficiently large constants. 

Conceptually speaking, the above description completely defines the
likelihood problem.  However, for the particular problem of
low-resolution analysis with high signal-to-noise data there is one
particular issue that must be considered in greater detail in order to
produce robust results, and that is the relationship between a finite
pixelization, bandwidth limitation, and covariance matrix
regularization.

For the expansion in equation \ref{eq:signal_mat} to be valid,
$\ell_{\textrm{max}}$ must be chosen sufficiently large such that
there is negligible power beyond this multipole. At the same time it
must be smaller than the corresponding Nyquist frequency of the chosen
pixelization. For HEALPix\footnote{http://healpix.jpl.nasa.gov/} the
recommended upper limit on this value is $\ell_{\textrm{max}} =
2\,N_{\textrm{side}}$ (where $N_{\textrm{pix}} = 12
N_{\textrm{side}}^2$), but through various numerical techniques it can
be pushed to $\ell_{\textrm{max}} = 3\,N_{\textrm{side}}$. Beyond
this, one is almost assured to get nonsensical results, and one must
therefore make sure that the experimental beam suppresses all signal
beyond this value. (A good rule-of-thumb is that the beam FWHM must be
at least 2.5 times the pixel size. For example, at $N_{\textrm{side}}
= 16$ the pixel size is $3.6^{\circ}$, and the smallest beam supported
by this pixelization is $9^{\circ}$ FWHM.)

In a pure signal map $\mathbf{s}$, with a given bandwidth limit
$\ell_{\textrm{max}}$, there are $(\ell_{\textrm{max}}+1)^2$ real
modes. Enforcing the recommended Nyquist limit of $\ell_{\textrm{max}}
= 2\,N_{\textrm{side}}$, a given pixelization can thus maximally
resolve $\sim 4\,N_{\textrm{side}}^2$ independent modes, which is
considerably less than the number of pixels in the map,
$N_{\textrm{pix}} = 12 N_{\textrm{side}}^2$. By a simple mode counting
argument, it is clear that a signal-only covariance matrix necessarily
will be singular, and consequently, it must be regularized in some way
before inversion.

The solution used by the currently available \emph{WMAP} likelihood
code is simply to increase $\ell_{\textrm{max}}$ to
$4\,N_{\textrm{side}}$, and thereby go beyond the Nyquist frequency of
the pixelization.  While it is true that this does indeed make the
matrix invertible (since the number of projected modes is now greater
than the number of pixels), it is also mathematically arbitrary,
highly pixelization dependent, and not connected to the observed data.

A much more reliable approach is readily available by means of the
noise covariance matrix. By adding a small amount of white noise to
the data, which has $N_{\textrm{pix}}$ independent modes, the matrix
becomes invertible, and the data description is still accurate. A
stable and well defined algorithm for computing low-resolution
likelihoods from high-resolution data may be formulated as follows:
\begin{enumerate}
\item Choose some $\ell_{\textrm{max}}$ of interest.
\item Determine the smallest pixelization that comfortably supports
  this scale and the corresponding pixel size $\theta_{\textrm{pix}}$.
\item Smooth the data with a Gaussian beam of
  $2.5\theta_{\textrm{pix}}$ FWHM.
\item Add Gaussian noise to the data with a variance such that the
  data are strongly noise dominated at $3N_{\textrm{side}}$.
\item Compute the likelihood as described above, with appropriately
  defined covariance matrices. 
\end{enumerate}

\section{B. Some Gibbs sampling extensions}
\label{app:powspec_methods}

Power spectrum estimation through Gibbs sampling was originally
introduced to CMB applications by \citet{jewell:2004} and
\citet{wandelt:2004}, and later implemented for high-resolution
applications like \emph{WMAP} by \citet{eriksen:2004} and
\citet{odwyer:2004}. The codes we use in the present paper are direct
descendants from those described in the latter two papers, but with a
few simple extensions we describe in this Appendix. Specifically, in
order to reduce the Markov chain correlation length in the low
signal-to-noise regime, we have implemented binned $C_{\ell}$ sampling
and sub-space sampling, and also sampling with Jeffrey's ignorance
prior which has an effect on large angular scales.

Let us first recall the basic idea of Gibbs sampling. Suppose we want
to draw samples from a complicated joint probability distribution
$P(x,y)$, but only know how to sample from the \emph{conditional}
distributions $P(x|y)$ and $P(y|x)$. Then the theory of Gibbs sampling
tells us that joint samples $(x^i, y^i)$ may be obtained by
alternately drawing from these two distributions, $x^{i+1} \leftarrow
P(x|y^i)$ and $y^{i+1} \leftarrow P(y|x^{i+1})$. As shown by
\citet{jewell:2004} and \citet{wandelt:2004}, this may be applied to
CMB analysis because it is in fact feasible to sample from both
$P(C_{\ell}|\mathbf{d}, \mathbf{s})$ and $P(\mathbf{s}|\mathbf{d},
C_{\ell})$ where $\mathbf{d}$ are observed data and $\mathbf{s}$ is
the true CMB sky. Explicitly, the algorithm may be described by these
two steps:


\begin{eqnarray}
\mathbf{s}^{i+1} &\leftarrow& P(\mathbf{s}|C_{\ell}^{i}, \mathbf{d}), \\
C_{\ell}^{i+1} &\leftarrow& P(C_{\ell}|\mathbf{s}^{i+1}).
\label{eq:gibbs}
\end{eqnarray}

Here $\leftarrow$ indicates sampling from the distribution on the
right. Thus, after some burn-in time $(C_{\ell}^i, \mathbf{s}^i)$ will
be drawn from the desired distribution, and these samples may
subsequently be used to establish marginal $P(C_{\ell}|\mathbf{d})$
and $P(\mathbf{s}|\mathbf{d})$ if desired. 

\subsection{Binned $C_{\ell}$ sampling}

One of the major problems with the implementations described by
\citet{eriksen:2004} was their inability to probe the low
signal-to-noise regime. The reason was simply that the Markov chain
step size between two consecutive $C_{\ell}$ samples is given by
cosmic variance alone, while the full posterior is given by both
cosmic variance and noise. Thus, in order to move a significant
distance in the high-$\ell$ regime, a very large number of steps is
required.

One way to improve on this situation is simply to bin the power
spectrum, and thereby increase the signal-to-noise ratio per sampled
parameter. (Note that such binning must be done internally in the
$P(C_{\ell}|\mathbf{d}, \mathbf{s}) \equiv P(C_{\ell}|\mathbf{s})$
sampler in order to earn an improvement -- binning by post-processing
does not have the same effect.)

Before describing the binned sampling algorithm, it is useful to
recall the single-$\ell$ algorithm for sampling from
$P(C_{\ell}|\mathbf{s})$: First compute the power spectrum of the
signal map, and write it for convenience on the form $\sigma_{\ell} =
\sum_{m=-\ell}^{\ell} |a_{\ell m}|^2$. Next, draw $2\ell-1$ Gaussian
random variates $\rho_{\ell}^j$ with zero mean and unit variance, and
form the sum $\rho_{\ell}^2 = \sum_{j=1}^{2\ell-1} |\rho_{\ell}^j|^2$.
Then the desired power spectrum sample is given by
\begin{equation}
C_{\ell}^{i+1} = \frac{\sigma_{\ell}}{\rho_{\ell}^2}
\end{equation}

Sampling binned $C_{\ell}$'s is very similar. First, we define the
binning scheme to be uniform in $C_{\ell}\,\ell(\ell+1)$, and choose
some bin limits $\ell_{\textrm{min}}$ and $\ell_{\textrm{max}}$. We
then form the quantity
\begin{equation}
  \sigma_{\textrm{b}} =
  \sum_{\ell=\ell_{\textrm{min}}}^{\ell_{\textrm{max}}}
  \sum_{m=-\ell}^{\ell} \ell(\ell+1)\,|a_{\ell m}|^2.
\end{equation}
The number of independent harmonic modes in this sum is $n \equiv
(\ell_{\textrm{max}}+1)^2 - \ell_{\textrm{min}}^2$, and therefore we
draw $n$ Gaussian random variates $\rho^j$ with zero mean and unit
variance. Next, we form the sum
\begin{equation}
\rho_{\textrm{b}}^2 =
\sum_{j=1}^{n} |\rho^j|^2.
\end{equation}
The common bin sample value is then 
\begin{equation}
C_{\textrm{b}} =
\frac{\sigma_{\textrm{b}}}{\rho_{\textrm{b}}},
\end{equation}
and the actual power spectrum sample coefficients are
\begin{equation}
  C_{\ell} = C_{\textrm{b}} / \ell(\ell+1).
\end{equation}

\subsection{Sub-space sampling}

A second technique to speed up convergence is sub-space sampling. As
described above, Gibbs sampling simply means sampling one parameter at
a time while conditioning on all others. If beneficial, we may
therefore choose to sample only a few $C_{\ell}$'s and
$\sigma_{\ell}$'s at a time while conditioning on all others.

Specifically, we may extend the basic sampling scheme given in
Equation \ref{eq:gibbs} as follows.
\begin{eqnarray}
  \mathbf{s}_{\textrm{low}}^{i+1} &\leftarrow&
  P(\mathbf{s}_{\textrm{low}}|C_{\ell,\textrm{low}}^{i}, C_{\ell,\textrm{high}}^{i}, \mathbf{s}^{i}_{\textrm{high}}, \mathbf{d}), \\
  C_{\ell,\textrm{low}}^{i+1} &\leftarrow&
  P(C_{\ell,\textrm{low}}|\mathbf{s}^{i+1}_{\textrm{low}}), \\
  \mathbf{s}_{\textrm{high}}^{i+1} &\leftarrow&
  P(\mathbf{s}_{\textrm{high}}|C_{\ell,\textrm{high}}^{i}, C_{\ell,\textrm{low}}^{i},
  \mathbf{s}^{i+1}_{\textrm{low}}, \mathbf{d}), \\
  C_{\ell,\textrm{high}}^{i+1} &\leftarrow&
  P(C_{\ell,\textrm{high}}|\mathbf{s}^{i+1}_{\textrm{high}}).
\end{eqnarray}
Sampling from $P(C_{\ell}|\mathbf{s})$ for a sub-set follows exactly
the same algorithm as before. For
$P(\mathbf{s}_{\textrm{low}}|C_{\ell}, \mathbf{s}_{\textrm{high}},
\mathbf{d})$ two trivial modifications must be made: The complementary
sky signal that is conditioned upon must be subtracted from the data
prior to sampling, and the corresponding $C_{\ell}$ coefficients must
be set to zero.

The advantage of this partitioning lies in the relationship between
pre-conditioning performance and correlation length: The Markov chain
correlation length is very short in the high signal-to-noise regime
but very long in the low signal-to-noise regime. Thus, in principle we
would like to make a larger number of steps at high $\ell$'s than at
low $\ell$'s, in order to obtain similar mixing properties
everywhere. On the other hand, most of the computational expense for
Gibbs sampling is spent on sampling from $P(\mathbf{s}|C_{\ell},
\mathbf{d})$ for which a linear system must be solved using Conjugate
Gradients. This linear system is dense in the high signal-to-noise
regime, but nearly diagonal in the low signal-to-noise
regime. Therefore, by conditioning on the computationally expensive
high signal-to-noise components, we can sample the high-$\ell$
components more aggressively with a low computational cost per sample.

For the analysis presented here, this is implemented through the
following sampling scheme:
\begin{itemize}
\item For each main sample stored on disk,
  \begin{itemize}
  \item draw one all-scale sample ($\sim120$ CG iterations).
  \item draw three intermediate-scale ($\ell \ge 300$) samples
  ($\sim20$ CG iterations).
  \begin{itemize}
  \item draw three small-scale ($\ell \ge 400$) samples ($\sim7$ CG
  iterations) for each intermediate-scale sample.
  \end{itemize}
\end{itemize}
\end{itemize}
The pre-conditioner may be individually tuned to each case. In
particular, an expensive pre-conditioner is used for the first case,
and a cheap, diagonal pre-conditioner is used for the latter
two. Also, if for a particular application one finds that convergence
is slow for low $\ell$'s (such as foreground sampling), one may
condition on, say, $\ell > 50$ and solve the system exactly in one
single iteration using the pre-conditioner described by
\citet{eriksen:2004}.

To summarize, it is straightforward to obtain good convergence even in
the low signal-to-noise regime using Gibbs sampling with the
introduction of binned $C_{\ell}$ sampling and sub-space sampling. For
the V-band analyses presented in this paper ($N_{\textrm{side}}=512$,
$\ell_{\textrm{max}}=700$), we produced 4000 such de-correlated
samples in three days with 16 processors, reaching a Gelman-Rubin
statistic value of $R-1 < 0.05$ for the last bin and $R-1 < 0.01$ for
$\ell < 600$.

\subsection{Jeffrey's prior}

The Bayesian approaches to power spectrum estimation require an
explicit statement of the prior probability distribution of the power
spectrum.  This prior reflects the experimenter's knowledge, or lack
thereof, about the power spectrum before the data are collected.

The Jeffreys prior is often useful because it encodes a complete lack
of knowledge about scale.  If we have some parameter, such as a value
of $C_\ell$, which we know is positive, but have no idea of its order
of magnitude, then we can use $P(C_\ell) = 1 / C_\ell$ as our prior.
This is the Jeffreys prior, and it gives equal weight to each
logarithmic bin, reflecting the initial belief that $C_\ell$ is
equally likely to be in any of them.

In practice, the scale of the parameter in question is not completely
unknown.  In the case of the CMB, for example, $\sqrt{C_\ell} <
2.7\mbox{K}$, since the temperature of the CMB cannot be negative
anywhere on the sky.  While this information should technically be
included in the prior, it is often not necessary to include it.  In
this case, the data already constrain the values of $C_\ell$ so
strongly that the above cutoff in the prior would have no effect on
the final posterior distribution.

The Jeffreys prior weights the posterior probability more toward low
values of $C_\ell$ than a uniform prior would.  The uniform prior
($P(C_\ell) = \mbox{constant}$) is useful because the posterior is
then exactly equal to the likelihood.  We plot the posterior with both
priors in figure 4, to show the effect of the Jeffreys prior in the
final posterior distribution.

\end{document}